# A Probabilistic Model For Sequence Analysis

Amrita Priyam
Dept. of Computer Science and Engineering
Birla Institute of Technology
Ranchi, India.

B. M. Karan[+], G. Sahoo[++]
[+]Dept. of Electronics and Electrical Engineering
[++]Dept. of Information Technology
Birla Institute of Technology
Ranchi, India

*Abstract—* **This paper presents a probabilistic approach for DNA sequence analysis. A DNA sequence consists of an arrangement of the four nucleotides A, C, T and G and different representation schemes are presented according to a probability measure associated with them. There are different ways that probability can be associated with the DNA sequence: one way is when the probability of an occurrence of a letter does not depend on the previous one (termed as unsuccessful probability) and in another scheme the probability of occurrence of a letter depends on its previous letter (termed as successive probability). Further, based on these probability measures graphical representations of the schemes are also presented. Using the digram probability measure one can easily calculate an associated probability measure which can serve as a parameter to check how close is a new sequence to already existing ones.**

**Keywords-Successive Probability; Unsuccessive Probability; Transition Probability; Digram Probability;**

## I. INTRODUCTION

A DNA sequence is a succession of the letters A, C, T and G. The sequences are any combination of these letters. A physical or mathematical model of a system produces a sequence of symbols according to a certain probability associated with them. This is known as a stochastic process, that is, it is a mathematical model for a biological system which is governed by a set of probability measure. The occurrence of the letters can lead us to the further study of genetic disorder. The stochastic process is also known mathematically as Discrete Markov Process. There are different ways to use probabilities for depicting the DNA sequences. One scheme is where each occurrence of a letter is independent of the occurrence of any other letter. That is, the probability of occurrence of a letter does not depend on the occurrence of the previous one. In yet another scheme, the occurrence of a letter depends on the occurrence of the previous one. The earlier one is termed in as unsuccessive probability and the later one is termed as successive probability. From this study we can further show the relationship between sequence and genetic variations. It can also lead to a more powerful test for identifying particular classes of genes or proteins which has been illustrated by an example.

## II. DNA SEQUENCE

DNA sequences is a succession of letters representing the primary structure of a real or hypothetical DNA molecule or strand, with the capacity to carry information as described by the central dogma of molecular biology. There are 4 nucleotide bases (A – Adenine, C – Cytosine, G – Guanine, T – Thymine). DNA sequencing is the process of determining the exact order of the bases A, T, C and G in a piece of DNA. In essence, the DNA is used as a template to generate a set of fragments that differ in length from each other by a single base. The fragments are then separated by size, and the bases at the end are identified, recreating the original sequence of the DNA. The most commonly used method of sequencing DNA the dideoxy or chain termination method was developed by Fred Sanger in 1977 (for which he won his second Nobel Prize). The key to the method is the use of modified bases called dideoxy bases; when a piece of DNA is being replicated and a dideoxy base is incorporated into the new chain, it stops the replication reaction.

Most DNA sequencing is carried out using the chain termination method. This involves the synthesis of new DNA strands on a single standard template and the random incorporation of chain-terminating nucleotide analogues. The chain termination method produces a set of DNA molecules differing in length by one nucleotide. The last base in each molecule can be identified by way of a unique label. Separation of these DNA molecules according to size places them in correct order to read off the sequence.

## III. DIFFERENT PROBABILISTIC APPROACHES FOR SEQUENCE REPRESENTATION

A DNA sequence is essentially represented as a string of four characters A, C, T, G and looks something like ACCTGACCTTACG. These strings can also be represented in terms of some probability measures and using these measures it can depicted graphically as well. This graphical representation matches the Markov Hidden Model. Some of these schemes are presented in this paper. A physical or mathematical model of a system produces a sequence of symbols according to a certain probability associated with them. This is known as a stochastic process. There are different ways to use probabilities for depicting the DNA sequences

A. *Unsuccessive Probability*

In this representation scheme the probability of the next occurrence of a letter does not depend on the previous letter. There can be two different representation schemes in this: one which uses equal probability for each letter and another which assumes a fixed probability for each letter.

- *Equal Probability* - Suppose we have a 4 letter code consisting of the 4 letters A, C, T, G which can be chosen each with equal probability 0.25, successive choices being independent. This would lead to a sequence of which the following is a typical example

  ACCATGGACTTAGCTACTGG

- *Unequal Probability* - For a DNA sequence series of length 20, where each letter has a probability of 0.3, 0.2, 0.3, 0.2 respectively, with successive choices are independent. A typical message from this source is:

  ACTTGAAATTCGGACCTGAT

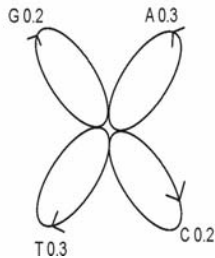

*Fig 1: Graphical Representation of Unequal Probability Sequence*

B. *Successive Probability*

This section presents two schemes for representing a DNA sequence: one where successive letter depends on the preceding one and in another scheme the letters are used to form words and a probability is associated with each word.

- *Letter Probability* - A more complicated structure is obtained if successive symbols are not chosen independently but their probabilities depend on preceding letters. In the simplest case of this type a choice depends only on the preceding letter and not on ones before that. The statistical structure can then be described by a set of transition probabilities $p_i(j)$, the probability that a letter $i$ is followed by letter $j$. The indices $i$ and $j$ range over all the possible symbols. A second equivalent way of specifying the structure is to give the "digram" probabilities $p(i,j)$, i.e., the relative frequency of the digram $i\ j$. The letter frequencies $p(i)$, (the probability of letter $i$), the transition probabilities $p_i(j)$ and the digram probabilities $p(i,j)$ are related by the following formulas:

$$p(i) = \sum_j p(i,j) = \sum_j p(j,i) = \sum_j p(j)p_j$$

$$p(i,j) = p(i)p_i(j)$$

$$\sum_j p_i(j) = \sum_i p(i) = \sum_{i,j} p(i,j) = 1$$

In the example presented in this paper we have used the letter probability for representing the sequence.

- *Word Probability* - A process can also be defined which produces a text consisting of a sequence of words. Since a DNA sequence consists of typically 4 letters A, C, T, G any word could be made up of these four letters. For example the typical DNA sequence may consist of any combination of the following words:
  ACTG TACG ACGT AATC AGTG TCCA CAAG CCTG

  This can be depicted graphically as in figure 2.

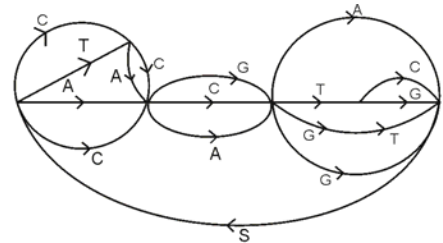

*Fig 2: Graphical Representation of Word Probability Sequence*

The various probability (transition and digram) are calculated according to the algorithm given below:

**ALGORITHM**

SERIES (DATA, N, LOC): Here DATA is a linear array with N elements and LOC acts as a pointer that keeps the record of occurrence of the nucleotide.

1. Initialize LOC = 0  and initialize all variables to zero.

2. Repeat while LOC ≠ N,
       IF DATA [LOC] = 'A'
       Set    A = A+ 1
       IF DATA [LOC +1 ]  = 'A' SET AA = AA+1

```
IF DATA [LOC +1 ] = 'C'  SET AC = AC+1
IF DATA [LOC +1 ] = 'T'  SET AT = AT+1
IF DATA [LOC +1 ] = 'G'  SET AG = AG +1

Else IF DATA [LOC] = 'C'
 Set  C = C+1
IF DATA [ LOC + 1] = 'A'  SET CA = CA+1
IF DATA [LOC +1 ] = 'C'  SET CC = CC+1
IF DATA [LOC +1 ] = 'T'  SET CT = CT+1
IF DATA [LOC +1 ] = 'G'   SET CG = CG +1

Else IF DATA [LOC] = 'T'
Set T = T+1
IF [LOC +1 ] = 'A'     SET TA = TA+ 1
IF [LOC +1 ] = 'C'     SET TC = TC+1
IF [LOC + 1 ] = 'T'    SET TT = TT + 1
IF [LOC + 1 ] = 'G'    SET TG = TG +1

Else IF DATA [LOC] = 'G'
Set G = G+1
IF DATA [LOC +1] = 'A'   SET GA = GA +1
  IF DATA [LOC +1] = 'C'   SET GC = GC +1
  IF DATA [LOC +1] = 'T'   SET GT = GT +1
  IF DATA [LOC+1] = 'G'    SET GG = GG+1
```

3. SET A' = A/N , C' = C/N , G' = G/N,  T' = T / N
   AA' = AA* A';  AC' = AC*A'; AT ' = AT* A';
   AG ' =AG* A'; CA'  = CC* C' ; CC' = CC*C';
   CT' = CT * C'; CG' = CG * C'; TA' = TA * T';
   TC' = TC * T'; TT' =TT * T'; TG' = TG* T';
   GA' = GA *G'; GC' = GC *G'; GT' = GT * G';
   GG' = GG * G';

4. End

### LETTER FREQUENCY

| (i) | P(i) |
|---|---|
| A | No of A'S / Size of string = (A' Say) |
| C | No of C'S / Size of string = ( C' Say ) |
| T | No of T'S / Size of string  = ( T' Say ) |
| G | No of G'S / Size of string = (G' say ) |

### DIGRAM PROBABILITY

| P(i, j) | A | C | T | G |
|---|---|---|---|---|
| A | No. of AA * A' = AA' | No. of AC * A' = AC' | No. of AT * A' = AT' | No. of AG * A' = AG' |
| C | No. of CA * C' = CA' | No. of CC * C' = CC' | No. of CT * C' = CT' | No. of CG * C' = CG' |
| T | No. of TA * T' = TA' | No. of TC * C' = TC' | No. of TT * T' = TT' | No. of TG * T' = TG' |
| G | No. of GA * G' = GA' | No. of GC * C' = GC' | No. of GT * T' = GT' | No. of GG * G' = GG' |

### TRANSITION PROBABILITY

| $P_i(j)$ | A | C | T | G |
|---|---|---|---|---|
| A | A' * AA' | A' * AC' | A' * AT' | A' * AG' |
| C | C' * CA' | C' * CC' | C' * CT' | C' * CG' |
| T | T' * TA' | T' * TC' | T' * TT' | T' * TG' |
| G | G' * GA' | G' * GC' | G' * GT' | G' * GG' |

On using the algorithm on the different H1N1 viruses we get the following transition probability tables:

Type 1:

**TABLE I**

|   | A | C | T | G |
|---|---|---|---|---|
| A | 0.13 | 0.06 | 0.09 | 0.08 |
| C | 0.09 | 0.04 | 0.05 | 0.02 |
| T | 0.06 | 0.05 | 0.05 | 0.07 |
| G | 0.08 | 0.04 | 0.05 | 0.06 |

Type 2:

**TABLE II**

|   | A | C | T | G |
|---|---|---|---|---|
| A | 0.13 | 0.06 | 0.09 | 0.08 |
| C | 0.08 | 0.04 | 0.04 | 0.02 |
| T | 0.06 | 0.05 | 0.06 | 0.07 |
| G | 0.08 | 0.04 | 0.04 | 0.06 |

Type 3:

**TABLE III**

|   | A | C | T | G |
|---|---|---|---|---|
| A | 0.08 | 0.05 | 0.08 | 0.08 |
| C | 0.08 | 0.03 | 0.06 | 0.02 |
| T | 0.05 | 0.06 | 0.06 | 0.09 |
| G | 0.08 | 0.06 | 0.04 | 0.07 |

A graphical representation for the type 1 H1N1 virus can be constructed based on the digram probability as follows:

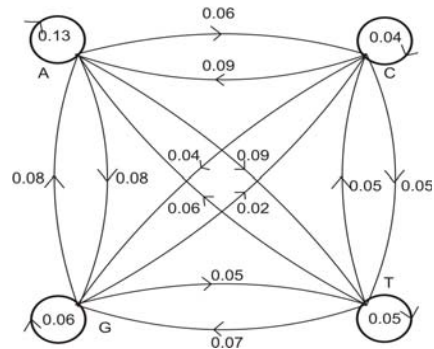

*Fig 3: Graphical Representation of Type 1 H1N1 Sequence*

Now if any new sequence comes which has to be categorized as any one of the three types we can use the transition probability table of the existing types and calculate the probability measure. For example consider a new sequence ***tggtgctggc***

Probability for Type 1: 0.07 * 0.06 * 0.05 * 0.07 * 0.04 * 0.05 * 0.07 * 0.06 * 0.04 = $4.9 \times 10^{-12}$

Probability for Type 2: 0.06 * 0.06 * 0.04 * 0.07 * 0.04 * 0.04 * 0.07 * 0.06 * 0.04 = $2.7 \times 10^{-11}$

Probability for Type 3: 0.09 * 0.07 * 0.04 * 0.09 * 0.06 * 0.06 * 0.09 * 0.07 * 0.06 = $3 \times 10^{-11}$

Since the probability measure for type 3 comes out to be the greatest we can conclude that the chance for the new DNA sample to be the type 3 virus is more.